\documentclass[11pt]{article}

\usepackage[utf8]{inputenc}
\usepackage[T1]{fontenc}
\usepackage[margin=1in]{geometry}
\usepackage{amsmath,amssymb}
\usepackage{graphicx}
\usepackage{booktabs}
\usepackage{multirow}
\usepackage{hyperref}
\usepackage{xcolor}
\usepackage{enumitem}
\usepackage[font=small]{caption}
\usepackage{url}
\usepackage{natbib}
\usepackage{microtype}
\usepackage{lineno}
\usepackage[ruled,vlined]{algorithm2e}

\newcommand{\keywords}[1]{\par\addvspace\medskipamount{\noindent
\textbf{\textit{Keywords}}\enspace\ignorespaces#1\par}}

\title{Learnable Pulse Accumulation for On-Device Speech Recognition:\\[4pt]
{\large How Much Attention Do You Need?}}

\author{Yakov Pyotr Shkolnikov \\
\texttt{yshkolni@gmail.com}}

\date{March 2026}

\begin{document}
\maketitle

\begin{abstract}
Self-attention scales quadratically with sequence length, limiting
transformer-based speech models on edge devices. We introduce the
Learnable Pulse Accumulator (LPA), an $O(n)$ replacement that
substitutes key-query dot products with learned gating functions:
content-dependent rectangular pulses, periodic windows, and
position-dependent basis functions. An MSE diagnostic sweep determines
per-layer replacement difficulty and ordering. Replacing 8 of 12
wav2vec2-base layers yields 10.61\% word error rate (WER) on
LibriSpeech test-clean, $+$7.24 percentage points~(pp) over the
3.37\% baseline, with 3.27$\times$ speedup at 120\,s audio on
Apple M4 Pro via an optimized MLX inference path. Cross-domain
validation on SepFormer speech enhancement shows all 16 intra-chunk
attention layers can be replaced without collapse, suggesting the
depth wall arises from linguistic computation rather than an LPA
limitation. LPA's near-binary gates at inference enable dense GPU
computation with no CPU--GPU synchronization, and all operations
map to mobile neural accelerators.
\end{abstract}


\section{Introduction}
\label{sec:intro}

Self-attention~\cite{vaswani2017attention} computes pairwise interactions
across all sequence positions, requiring $O(n^2 d)$ operations for
sequence length $n$ and hidden dimension $d$. In speech, 50\,Hz frame
rates produce thousands of frames per utterance. At these lengths,
attention dominates memory at all durations and latency beyond
$\sim$20\,s of audio.

On mobile hardware, the problem is compounded. Neural accelerators
such as Apple's ANE and Qualcomm Hexagon achieve peak throughput on element-wise
and convolution operations but lack efficient support for the dynamic
$n {\times} n$ matrix multiplications that attention requires. This
forces GPU fallback and negates the accelerator's power efficiency. An
ASR model whose mixing operations are entirely accelerator-compatible could
achieve an order-of-magnitude speedup on device.

\begin{figure}[t]
\centering
\includegraphics[width=\textwidth]{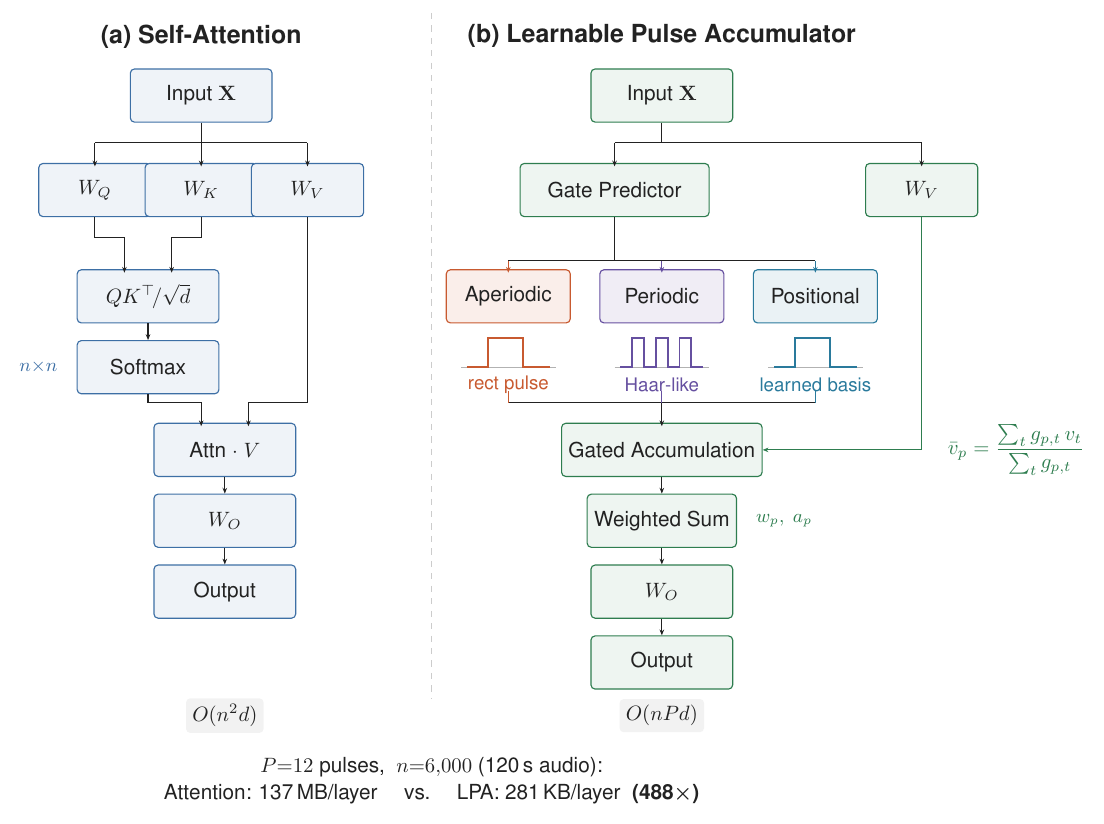}
\caption{(a) Standard self-attention computes an $n{\times}n$ matrix via
$QK^\top$. (b) LPA replaces this with three learned gate types that define
soft windows over the sequence. Gated accumulation produces per-pulse
summaries at $O(nP)$ cost. All operations are accelerator-compatible.}
\label{fig:architecture}
\end{figure}

We propose the \textbf{Learnable Pulse Accumulator (LPA)}, which
replaces key-query matching with learned gating functions that define
soft windows over the sequence. Three gate types capture complementary
patterns. \emph{Aperiodic} pulses handle content-dependent segmentation,
\emph{periodic} pulses for multi-scale rhythmic structure, and
\emph{positional} pulses for fixed structural biases. The computation
uses only depthwise convolution, sigmoid gating, and weighted
summation, all of which are accelerator-compatible operations
(Fig.~\ref{fig:architecture}).

LPA is architecturally distinct from prior $O(n)$ alternatives. Linear
attention~\cite{katharopoulos2020transformers} approximates the softmax
kernel. State space models (SSMs)~\cite{gu2023mamba} use recurrent state
evolution. SummaryMixing~\cite{summarymixing2024} uses mean pooling.
LPA instead uses \emph{explicit learned windows} where each pulse
defines where to aggregate information, how wide to look, and at what
periodicity to repeat.

We convert a pretrained wav2vec2-base model to LPA via progressive
layer-by-layer replacement, developing the techniques essential for
preserving quality. Our contributions:
\begin{enumerate}[nosep,leftmargin=*]
\item An $O(n)$ primitive, gated pulse accumulation, with three
complementary gate types for sequence mixing.
\item An MSE diagnostic sweep that measures per-layer replacement
difficulty and determines optimal replacement order, providing a
practical recipe for converting any pretrained transformer.
\item A roofline analysis demonstrating that pulse count has negligible
impact on single-sample inference because additional pulses on hard layers
add negligible cost.
\item Inference benchmarks on consumer hardware showing 3.27$\times$
speedup at 120\,s via a fused MLX/Metal path, with linear
scaling and full mobile accelerator compatibility.
\end{enumerate}

\section{Related work}
\label{sec:related}

\textbf{Efficient attention and attention-free models.}
Linear attention~\cite{katharopoulos2020transformers} replaces the
softmax kernel with a feature map ($O(nd^2)$). Gated Linear
Attention~\cite{yang2024gated} adds data-dependent forget gates with
hardware-efficient chunkwise training.
FlashAttention~\cite{dao2022flashattention} optimizes IO but retains
$O(n^2)$ complexity. The Attention Free Transformer
(AFT)~\cite{zhai2021aft} replaces dot-product attention with
element-wise sigmoid gating and learned position biases, but its
full formulation retains $T{\times}T$ position parameters and does not
use explicit windowed accumulation. Mega~\cite{ma2023mega} combines
exponential moving average with single-head gated attention, achieving
linear complexity on long-range benchmarks. The Learnable Multi-Scale
Wavelet Transformer~\cite{kiruluta2025lmwt} replaces dot-product
attention with a learned Haar wavelet decomposition, achieving linear
scaling on machine translation. These primarily target
language modeling or general sequence tasks.

\textbf{Efficient speech models.}
SummaryMixing~\cite{summarymixing2024} replaces attention in wav2vec~2.0
with mean-pooling branches (18\% faster, no quality drop on downstream
tasks). The Polynomial Mixer~\cite{pom2026} uses polynomial
representations, outperforming SummaryMixing on LibriSpeech~\cite{panayotov2015librispeech}. Fast
Conformer~\cite{fastconformer2023} uses 8$\times$ downsampling for
linear scaling (4.99\% WER, 2.8$\times$ speedup).
Zipformer~\cite{yao2024zipformer} uses a U-Net encoder with temporal
downsampling, achieving faster inference than Conformer.
LiteASR~\cite{kamahori2025liteasr} compresses Whisper via low-rank
approximation of activations (50\% encoder size reduction).
Mamba-based approaches
(ConMamba~\cite{speech_slytherin2024}, Samba-ASR~\cite{samba2025})
replace attention with SSMs but train from scratch on 10k+ hours.

\textbf{Progressive layer replacement.}
LoLCATs~\cite{lolcats2025} linearizes LLM attention via MSE distillation
and LoRA. Mamba-in-the-Llama~\cite{mamba_llama2024} replaces attention
with Mamba2 stepwise, finding end-to-end fine-tuning most impactful.
GatedDeltaNet~\cite{gated_deltanet2025} reports a 3:1
linear-to-attention ratio optimal in production (Qwen3-Next), consistent
with our finding that 8/12 (2:1) LPA layers is the practical limit.

LPA differs from AFT and Mega in its use of \emph{explicit square pulse
windows} (aperiodic, periodic, positional) rather than global
element-wise gating or exponential decay. Square pulses produce sparse,
near-binary gate activations after temperature annealing, which both
reduce arithmetic intensity and enable kernel fusion across consecutive
LPA layers (the sparsity pattern is determined by a few gate parameters
rather than a dense $T{\times}T$ bias). LPA differs from SummaryMixing
and LiteASR in replacing the mixing mechanism itself rather than using
pooling or low-rank compression. The combination of gated pulse
accumulation, speech/connectionist temporal classification (CTC) domain, and progressive replacement of
pretrained attention is, to our knowledge, unexplored.

\section{Learnable Pulse Accumulator}
\label{sec:lpa}

Given input $\mathbf{X} \in \mathbb{R}^{n \times d}$, LPA maintains $P$
pulses, each with a gate $g_p \in [0,1]^n$ assigning soft membership
to each position. The output at position $t$ is
\begin{equation}
\text{LPA}(\mathbf{X})_t = W_O
\frac{\sum_{p} w_p \, g_{p,t} \, a_p \, \bar{\mathbf{v}}_p}
     {\sum_{p} w_p \, g_{p,t}}
\label{eq:lpa}
\end{equation}
where $w_p$ are softmax-normalized pulse weights, $a_p$ are per-pulse
amplitudes, $W_O \in \mathbb{R}^{d \times d}$ is the output projection,
and $\bar{\mathbf{v}}_p$
is the gated mean of value-projected hidden states within pulse $p$.
An active mask $m_t = 1 - \exp(-\sum_p g_{p,t})$ suppresses output at
positions with no gate coverage, applied as a multiplicative factor on
the output:
\begin{equation}
\bar{\mathbf{v}}_p = \frac{\textstyle\sum_t g_{p,t} \cdot W_V \mathbf{x}_t}
                          {\textstyle\sum_t g_{p,t}},
\quad W_V \in \mathbb{R}^{d \times d}
\end{equation}

\subsection{Gate types}

\textbf{Aperiodic gates} predict center $c_p$ and half-width
$\delta_p$ from the input via a learned query attention mechanism.
\begin{equation}
g_p^{\text{a}}(t) = \sigma\!\Big(\tfrac{t - c_p + \delta_p}{\tau}\Big)
\cdot \sigma\!\Big(\tfrac{c_p + \delta_p - t}{\tau}\Big)
\end{equation}
where $\sigma(\cdot)$ denotes the sigmoid function,
$c_p = \sum_t t \cdot \text{softmax}(\mathbf{h}^\top
\mathbf{q}_p / \tau)_t$ with
$\mathbf{h} = \text{MLP}(\text{DWConv}(\mathbf{X})) \in \mathbb{R}^{n \times d/2}$
(MLP = multi-layer perceptron, DWConv = depthwise convolution),
$\mathbf{q}_p \in \mathbb{R}^{d/2}$ is a learned query vector for
pulse $p$, and
$\delta_p = \text{softplus}(f(\bar{\mathbf{h}}_p)) > 0$ ensures a
valid interval, where $\bar{\mathbf{h}}_p$ is the attention-weighted
mean of $\mathbf{h}$ using the same softmax weights as $c_p$.
Temperature $\tau$ anneals from soft to hard during
training.

\textbf{Periodic gates} predict period $T_p$, phase $\phi_p$, and
duty cycle $d_p$.
\begin{equation}
g_p^{\text{per}}(t) = \sigma\!\Big(
\tfrac{\cos(2\pi t/T_p - \phi_p) - \cos(\pi d_p)}{\tau}\Big)
\end{equation}
Period is parameterized as $T_p = 2^{\text{softplus}(\cdot)+2}$ (minimum
4 frames), initialized across scales from $\sim$10 to 512 frames to capture
phoneme-to-word-level repetitions.

\textbf{Positional gates} are content-independent. Each gate is a
learned linear combination of $K$ sin/cos basis functions over
normalized position $\hat{t} = t/(n{-}1)$, passed through sigmoid:
$g_p^{\text{pos}}(t) = \sigma\!\big((\sum_{k=1}^{K} \alpha_{p,k}
\sin(2\pi k \hat{t}) + \beta_{p,k} \cos(2\pi k \hat{t}) + b_p) / \tau\big)$
where $\alpha_{p,k}, \beta_{p,k} \in \mathbb{R}$ are learned
coefficients and $b_p$ is a bias.
With $K{=}16$ bases, these can represent arbitrary positional
patterns including telegraph-like bi-level signals, capturing fixed
structural biases that complement the content-dependent gates.

\subsection{Architecture details}

The aperiodic gate predictor uses a causal depthwise convolution (kernel 5)
followed by a 2-layer MLP with GELU activation. Periodic and positional
predictors use a single linear projection. Cross-layer coordination adds
a projected summary of the previous LPA layer's mean gate pattern as
an additive bias. The value projection $W_V$ and output projection $W_O$
are initialized from the original attention weights.

\subsection{Complexity}

Per-layer cost is $O(nPd + nd^2)$ vs.\ attention's $O(n^2d + nd^2)$.
With $P{=}12$ and $n{=}6000$ (120s audio), gate memory is 281\,KB
vs.\ attention's 137\,MB, a $500\times$ reduction. At
batch size 1, the $nd^2$ projections dominate and the $nPd$
accumulation is memory-bandwidth-bound on the input tensor, making $P$
effectively free (Appendix~\ref{app:roofline}).

\section{Progressive replacement recipe}
\label{sec:training}

We replace attention layers in a pretrained wav2vec2-base-960h~\cite{baevski2020wav2vec}
(3.34\% WER on dev-clean) one at a time, progressing from easiest to
hardest. The full recipe (Algorithm~\ref{alg:recipe}) requires only
two inputs: the pretrained model and a training set.

\begin{algorithm}[t]
\caption{Progressive LPA replacement}\label{alg:recipe}
\DontPrintSemicolon
\SetKwInOut{Input}{Input}
\SetKwInOut{Output}{Output}
\Input{Pretrained model $\mathcal{M}$ with $L$ attention layers, training set $\mathcal{D}$, WER budget $B$}
\Output{Partially-replaced model with $k$ LPA layers}
\BlankLine
\textbf{Step 1: MSE diagnostic sweep}\;
\For{each layer $\ell = 0, \ldots, L{-}1$}{
  Replace layer $\ell$ with overprovisioned LPA ($P_{\mathrm{sweep}}$ pulses)\;
  Train 2 epochs: $\mathcal{L} = \mathrm{MSE}(\mathrm{LPA}(\mathbf{X}), \mathrm{Attn}(\mathbf{X})) + \lambda_1\|\mathbf{a}\|_1 + \lambda_2\|\mathbf{a}\|_2^2$\;
  Record MSE (difficulty) and surviving pulse count\;
  Restore original attention\;
}
Sort layers by increasing MSE $\to$ replacement order\;
\BlankLine
\textbf{Step 2: Progressive replacement}\;
\For{$i = 1, \ldots, L$ in sweep order}{
  Initialize LPA from attention weights ($W_O$, $W_V$, partial $W_Q$)\;
  MSE warm-start: 2 epochs against original attention output\;
  CTC training: 8 epochs with temperature annealing $\tau{:}\;3.0 \to 0.5$\;
  \quad Unfreeze FFN at $0.1\times$ lr, unfreeze layer norm\;
  Alignment: jointly fine-tune all LPA layers ($\le$5 epochs, $0.5\times$ lr)\;
  \quad Re-anneal all LPA temperatures globally\;
  \quad Auto-revert if WER increases\;
  \lIf{$\mathrm{WER} > B$}{\textbf{stop}}
}
Final joint fine-tuning: 8 epochs at $0.2\times$ lr\;
\end{algorithm}

\label{sec:mse_sweep}

The sweep (Step~1) produces both a difficulty ordering and a per-layer
pulse allocation, since elastic net regularization prunes unneeded
pulses on easy layers while retaining capacity on hard ones.

Three techniques in Step~2 proved critical. \emph{Selective
initialization} from attention weights and layer norm unfreezing
together account for $-31.8$\,pp (Table~\ref{tab:ablation}), the
single largest gain. \emph{FFN co-adaptation} (unfreezing at
$0.1\times$ lr) adds $-1.5$\,pp, since each layer's FFN was co-trained
with its original attention. \emph{Temperature curriculum}
contributes $-2.9$\,pp: during per-stage training only the current
layer anneals, while during alignment all LPA layers re-anneal
globally, providing smoother gradients for cross-layer adaptation.

\section{Experiments}
\label{sec:experiments}

\textbf{Model.} \texttt{facebook/wav2vec2-base-960h} (12 layers, 768
hidden, 95M params, 3.34\% WER on dev-clean), the standard pretrained
release fine-tuned with CTC on 960h of
LibriSpeech~\cite{panayotov2015librispeech}.

\textbf{Data.} LibriSpeech train-clean-100 (100h) for early ablations,
train-clean-360 (360h) for final configurations. Evaluation on
dev-clean (2,703 utterances), greedy CTC decoding without
language model.

\textbf{LPA config.} Base: 4 aperiodic + 4 periodic + 4 positional =
12 pulses per layer, kernel size 5, deep gate predictor (2-layer MLP
for aperiodic, single projection for periodic/positional),
cross-layer coordination, value projection, output gate, 4 heads,
position modulation, dynamic content-dependent pulse weights, and skip
connections from 4~layers back. In the +Order configuration,
sweep-derived allocations override pulse counts (up to 36 for hard
layers). The +Architecture configuration uses a fixed override of
8+8+8=24 pulses for deep stages ($\ge$9).

\textbf{Training.} 8 epochs/stage at lr $5{\times}10^{-4}$ with
10\% linear warmup (FFN at $5{\times}10^{-5}$), up to 5 alignment
epochs at $2.5{\times}10^{-4}$ with auto-revert, 8 final epochs at
$10^{-4}$. CTC loss uses mean reduction and infinite-loss zeroing.
MSE sweep: 2 epochs/layer with $\lambda_1{=}0.01$, $\lambda_2{=}0.001$,
$\theta{=}0.1$, floor $f{=}4$.
AdamW, BF16 mixed precision, batch 48 on NVIDIA H100.
Total training cost for the full 10-stage progressive replacement
(including MSE sweep, per-stage training, alignment, and final joint
tuning) is $\sim$4 GPU-hours on H100.

\subsection{Ablation}

Table~\ref{tab:ablation} isolates each technique at 8/12 layers,
where differences are most pronounced.

\begin{table}[t]
\centering
\caption{Cumulative ablation at 8/12 layers replaced (dev-clean WER~\%).
Each row adds one technique to the previous configuration.
Rows marked $\dagger$ are estimated by correcting an early evaluation bug
($-$9.3\,pp uniform offset, Appendix~\ref{app:eval_bug}). The bottom
three rows use the corrected evaluation directly.}
\label{tab:ablation}
\begin{tabular}{@{}lcc@{}}
\toprule
Configuration & WER & $\Delta$ \\
\midrule
Na\"ive replacement$^\dagger$ & 58.33 & --- \\
\quad + Selective init, unfreeze LN$^\dagger$ & 26.50 & $-$31.8 \\
\quad + Inter-stage alignment$^\dagger$ & 16.18 & $-$10.3 \\
\quad + FFN co-adapt (0.1$\times$ lr)$^\dagger$ & 14.64 & $-$1.5 \\
\quad + Temperature curriculum$^\dagger$ & 11.72 & $-$2.9 \\
\quad + Periodic + positional gates & 10.25 & $-$1.5 \\
\quad + Skip connections, multi-head, 360h data & 9.77 & $-$0.5 \\
\quad + MSE-ordered replacement & \textbf{9.35} & $-$0.4 \\
\bottomrule
\end{tabular}
\end{table}

\textbf{Negative results.} Doubling pulse count (8$\to$16) and
adding per-pulse width predictors provided no improvement because gate
capacity is not the bottleneck. Restricting temperature annealing to
only the newly-replaced layer (rather than globally) degraded
performance by 3--5\,pp, confirming the curriculum effect.

\subsection{Replacement order matters more than method}
\label{sec:order_matters}

The MSE sweep shows large variation in replacement difficulty:
early layers (0--2) have MSE 50--60$\times$ lower than the final
layer, with a general acoustic-to-linguistic trend
(Table~\ref{tab:mse_difficulty} in Appendix).
Fig.~\ref{fig:progressive_wer} shows the progressive WER trajectory
across four configurations that cumulatively add training recipe
improvements, architectural changes, and difficulty-informed ordering.
The most controlled comparison (``+Architecture'' vs.\ ``+Order'')
uses identical architectures, data (360h), and hyperparameters,
differing only in layer replacement order. Deferring the hard
layer~8 from position~3 to position~7 reduces 6/12 WER from
7.32\% to 5.64\% ($-$1.68\,pp, 23\% relative).

\begin{figure}[t]
\centering
\includegraphics[width=0.7\textwidth]{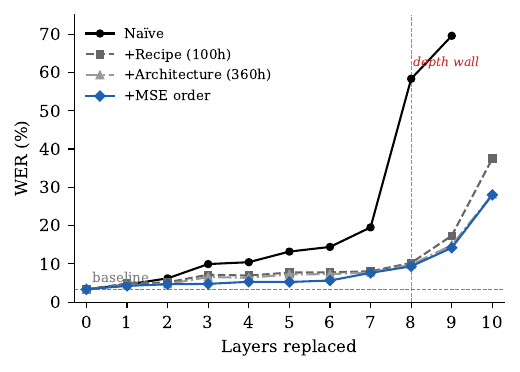}
\caption{WER (\%) vs.\ number of replaced layers across four
cumulative configurations
(full data in Table~\ref{tab:progressive}, Appendix).
MSE-ordered replacement yields lower WER at every stage.}
\label{fig:progressive_wer}
\end{figure}

At 8/12, WER drops from 58.33\% (na\"ive) to 9.35\% (full recipe)
on dev-clean, with 10.61\% on test-clean and 27.10\% on test-other
(Table~\ref{tab:test_splits}, Appendix). Test-clean tracks dev-clean
within $\sim$1\,pp at all depths.

\subsection{The depth wall}

All configurations degrade sharply beyond 8--9 replaced layers. The
transition from 9/12 to 10/12 adds $+$13--20\,pp depending
on configuration. The MSE sweep explains this directly. The
final layers are far harder to approximate. Layer~11 has
MSE 60$\times$ higher than layer~0 (0.370 vs.\ 0.006), and layer~10
is 15$\times$ harder.

We hypothesize this reflects a transition from acoustic to linguistic
computation across encoder depth (Sec.~\ref{sec:discussion}). The MSE
difficulty trend broadly tracks this transition, though
non-monotonically (layer~4 is harder than layers 7--9). For
deployment, the wall defines a quality--speed tradeoff curve.
The MSE sweep tells the practitioner which layers to replace and when
to stop.

\subsection{Cross-domain validation: speech enhancement}
\label{sec:sepformer}

To test whether the depth wall reflects linguistic computation
requirements (as hypothesized) rather than an LPA architectural
limitation, we apply the same progressive replacement recipe to
SepFormer~\cite{subakan2021sepformer}, a purely acoustic model for
speech enhancement.

\textbf{Setup.} SepFormer-WHAM16k ($\sim$26M params) uses a
dual-path architecture with $2 \times 8 = 16$ intra-chunk
and $2 \times 8 = 16$ inter-chunk attention layers
(embed\_dim=256, 8 heads, chunk size $K{=}250$).
We replace the 16 \emph{intra-chunk} layers progressively, using
MSE-ordered replacement. The 16 inter-chunk (global) attention
layers are retained. LPA uses 12 pulses per layer. Training and
evaluation use LibriSpeech audio mixed with Gaussian noise at
5\,dB SNR, with \emph{separate} train (dev-clean, 2703 files) and
held-out evaluation (test-clean, 2620 files) sets.

\textbf{Results.}
Table~\ref{tab:sepformer} shows scale-invariant signal-to-noise
ratio improvement (SI-SNRi, in dB) on the held-out set
during progressive replacement.

\begin{table}[t]
\centering
\caption{SepFormer SI-SNRi (dB) on held-out test-clean during
progressive LPA replacement of intra-chunk attention. All 16 layers
are replaced without collapse.}
\label{tab:sepformer}
\begin{tabular}{@{}rcc@{}}
\toprule
Layers & SI-SNRi & $\Delta$ from baseline \\
\midrule
0 (baseline) & 5.69 & --- \\
2/16 & 8.89 & $+$3.20 \\
4/16 & 8.92 & $+$3.23 \\
8/16 & 8.14 & $+$2.45 \\
11/16 & 7.68 & $+$1.99 \\
16/16 & 6.82 & $+$1.13 \\
\bottomrule
\end{tabular}
\end{table}

\textbf{No depth wall appears.} All 16
layers are replaced, with every stage remaining above
baseline. Performance peaks at 8.92 dB (4/16) and declines gently to
6.82 dB (16/16, still $+$1.13 above baseline), whereas
wav2vec2 collapses sharply at 8--9 layers. The MSE difficulty ratio
across SepFormer layers is $15\times$ (vs.\ $60\times$ in wav2vec2),
and difficulty does not correlate with depth, consistent with all layers
performing acoustic rather than linguistic computation.

\textbf{Fine-tuning confound.} The improvement above baseline (5.69
$\to$ 8.92~dB) is unexpected, since replacing attention with a
lower-capacity mechanism should not improve performance. A control experiment that
fine-tunes all 16 layers \emph{with attention intact} using the same
training budget reaches 11.34~dB, confirming that the improvement is
primarily a fine-tuning effect (the pretrained checkpoint was trained
on WHAM! noise, not Gaussian noise). The relevant comparison is
therefore LPA (6.82~dB at 16/16) vs.\ fine-tuned attention
(11.34~dB). LPA underperforms attention but does not collapse,
which is the depth-wall finding.

\textbf{Caveats.} The evaluation uses Gaussian noise (simpler than
WHAM!). SepFormer's smaller embedding (256 vs.\ 768) means FFN layers
dominate compute, so overall model speedup is only $\sim$1.23$\times$.
The value of the SepFormer experiment is the depth-wall validation,
not raw speedup.

\subsection{Inference speed}

\begin{table}[t]
\centering
\caption{Inference time (ms) on Apple M4 Pro, batch size 1.
\emph{PyTorch}: MPS backend.
\emph{MLX}: Metal GPU, FP16 with FP32 accumulation.
FP16 attention is slower than FP32 on MPS (the $n{\times}n$
matmul does not benefit from half precision on this hardware),
confirming that the MLX speedup reflects LPA's algorithmic advantage.}
\label{tab:speed}
\begin{tabular}{@{}rcccccc@{}}
\toprule
Audio & Base & Base & 8/12 & 12/12 & 8/12 & Speedup \\
      & (fp32) & (fp16) & (fp32) & (fp32) & (MLX) & vs fp16 \\
\midrule
10s  &   47 &   45 &   59 &   65 &   42 & 1.07$\times$ \\
30s  &  183 &  186 &  157 &  135 &  125 & 1.49$\times$ \\
60s  &  500 &  522 &  350 &  246 &  244 & 2.14$\times$ \\
120s & 1668 & 1767 &  895 &  479 &  540 & 3.27$\times$ \\
\bottomrule
\end{tabular}
\end{table}

In the PyTorch/MPS reference implementation, the crossover where LPA
becomes faster than attention occurs at approximately 20s, because
per-kernel dispatch overhead exceeds the quadratic cost saved at
shorter durations. A dedicated MLX inference path
(Sec.~\ref{sec:edge}), which computes all gate operations as dense GPU
kernels in half precision (fp16) and avoids CPU--GPU
synchronization, shifts this
crossover to below 10\,s. At 10\,s, MLX LPA achieves
1.07$\times$ speedup over the FP16 attention baseline.

\begin{figure}[t]
\centering
\includegraphics[width=0.7\textwidth]{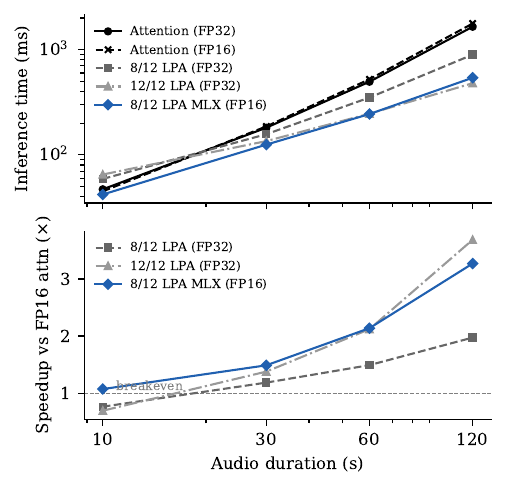}
\caption{Speedup vs.\ FP16 attention baseline on Apple M4 Pro, batch~1.
Top: inference time (ms, log scale). Bottom: speedup ratio.
FP16 attention is slower than FP32 on MPS (Table~\ref{tab:speed}).}
\label{fig:speedup_scaling}
\end{figure}

Speedup scales linearly with input length
(Fig.~\ref{fig:speedup_scaling}) because the $O(n)$ vs.\ $O(n^2)$ gap
widens indefinitely. At 120s, the real-time factor is 0.0045. At batch
size 1, linear projections dominate LPA time, making pulse count
effectively free: 12$\to$36 pulses adds only 0.6\%
(Appendix~\ref{app:roofline}). All LPA operations (depthwise convolution,
sigmoid, element-wise multiply, linear projection) are compatible with
mobile neural accelerators. FlashAttention is unavailable on Apple
MPS/Metal, so the $O(n^2)$ baseline reflects the actual on-device cost.

\section{Comparisons}
\label{sec:analysis}

\textbf{Comparison to SummaryMixing.}
SummaryMixing~\cite{summarymixing2024} replaces attention with
mean-pooled summaries: $\mathbf{s} = \text{Linear}(\text{mean}(\mathbf{X}))$,
broadcast and added to a local per-position projection.
Both LPA and SummaryMixing are $O(n \cdot d^2)$, but LPA learns
\emph{temporal selectivity} (which positions contribute to each output)
while SummaryMixing uses uniform averaging.

Table~\ref{tab:summarymixing} compares progressive replacement.
On SepFormer, both methods use identical training budgets. On wav2vec2,
the SummaryMixing implementation processes 2k samples per epoch
(vs.\ LPA's full $\sim$28k), giving LPA a training advantage that
prevents direct comparison. Despite this caveat, the directional
pattern is informative. LPA outperforms SummaryMixing at every wav2vec2
depth, with the gap widening from $+$1.2\,pp at 1/12 to $+$9.1\,pp
at 4/12, while on SepFormer (where budgets are matched) SummaryMixing
\emph{matches or exceeds} LPA at all depths. This asymmetry is
consistent with intra-chunk attention performing only local acoustic
smoothing that mean pooling captures well, while wav2vec2's linguistic
layers require temporal selectivity.

\begin{table}[t]
\centering
\caption{LPA vs.\ SummaryMixing (SM) progressive replacement.
wav2vec2: dev-clean WER (\%, $\downarrow$).
SepFormer: held-out test-clean SI-SNRi (dB, $\uparrow$, 16 intra-chunk layers).
SepFormer uses identical training budgets; wav2vec2 SM uses fewer
samples per epoch (see text).}
\label{tab:summarymixing}
\begin{tabular}{@{}rccrcc@{}}
\toprule
& \multicolumn{2}{c}{wav2vec2 WER (\%)} && \multicolumn{2}{c}{SepFormer SI-SNRi (dB)} \\
\cmidrule{2-3} \cmidrule{5-6}
Layers & LPA & SM && LPA & SM \\
\midrule
0 (base) & 3.18 & 3.18 && 5.69 & 5.69 \\
1  & 4.26 & 5.44 && 7.27 & 8.38 \\
2  & 4.64 & 7.31 && 8.89 & 9.60 \\
4  & 6.38 & 15.47 && 8.92 & 9.62 \\
8  & 9.92 & 29.04 && 8.14 & 9.12 \\
16 & --- & --- && 6.82 & 7.88 \\
\bottomrule
\end{tabular}
\end{table}

\textbf{Comparison to Mamba-based ASR.}
ConMamba~\cite{speech_slytherin2024} achieves 22.65\% WER with 80h data.
LPA at 8/12 achieves 10.61\% with 360h fine-tuning, though the
comparison is confounded by our pretrained wav2vec2 initialization
vs.\ training from scratch. CALD~\cite{cald2024} converts
Wav2Vec2-\emph{large} (317M params) attention to Mamba2 via layerwise
distillation, achieving $+$0.32\,pp WER on TED-LIUM. The comparison
is not direct because CALD uses a 3$\times$ larger model with more redundancy,
a different test set, and requires running the full attention model as
a teacher during conversion. LPA operates on the smaller wav2vec2-base
(95M params) without a teacher model, trading quality for a simpler,
teacher-free conversion pipeline. Both approaches confirm that post-hoc
attention conversion is viable for speech models.

\textbf{Quality gap in context.}
At 8/12 replacement, LPA adds $+$7.24\,pp over the 3.37\% baseline,
a meaningful degradation. Three factors suggest this gap is reducible
rather than fundamental.
First, the current recipe uses no distillation: each LPA layer is
trained with CTC loss alone, while CALD's near-lossless result
($+$0.32\,pp) and LoLCATs~\cite{lolcats2025} both rely on per-layer
MSE supervision from the original attention output as an auxiliary
signal. Preliminary experiments with MSE auxiliary loss showed
directional improvement but were confounded by simultaneous changes
to replacement order.
Second, the 3.27$\times$ speedup comes with 4 attention layers intact.
GatedDeltaNet~\cite{gated_deltanet2025} independently finds that a 3:1
linear-to-attention ratio is optimal in production (Qwen3-Next),
suggesting that retaining a minority of attention layers for
hard linguistic computation may be the right operating point rather
than a limitation.
Third, our hyperparameter search was limited: a single temperature
schedule ($\tau{=}3{\to}0.5$), fixed learning rates, and no
per-layer tuning of pulse allocation. Cross-layer skip connections
(which proved important for depth $\ge$9) and per-head gate
specialization were tested only in the final configuration. Systematic
tuning of these choices remains unexplored.

\section{Discussion: a new building block}
\label{sec:discussion}

Attention computes a data-dependent weighted average: the softmax of
key-query dot products selects \emph{which} positions contribute and
\emph{how much}. This is a matched filter, optimal when the target
signal is buried in competing signals at unknown positions. But when
the signal is spatially localized and spectrally distinct from noise,
a simpler estimator suffices.

Consider a signal $x(t) = s(t) + \eta(t)$ where $s(t)$ is a target
signal nonzero only on some interval $[a, b]$ and $\eta(t)$ is noise.
The sufficient statistic for estimating $s$
is the integral $\int_a^b x(t)\,dt$, uniform averaging over the
support, with no weighting needed. The only information the estimator
requires is the \emph{support} $[a, b]$ itself. This is what
LPA's gate predictor learns: where the relevant signal lives
(center and half-width), not what it looks like (key-query match
scores). The gate predictor replaces content matching with
localization, and the rectangular window replaces softmax weighting
with uniform integration.

Speech satisfies the localization premise. Phonemes occupy contiguous
time intervals. Coarticulation effects span at most a few hundred
milliseconds. The acoustic features at layers~0--7 of wav2vec2
are dominated by local spectro-temporal structure, the
regime where gated integration extracts a sufficient statistic
and the matched-filter machinery of attention is unnecessary. The MSE
difficulty gradient quantifies this. Easy layers
(MSE~$\sim$0.006) have attention patterns that are already
near-rectangular, while hard layers (MSE~$\sim$0.37) have the
content-dependent routing that rectangular windows cannot capture.

The three gate types cover complementary points on the Gabor
time-frequency tradeoff~\cite{bao2022reganet,wang2025hybornet}:
aperiodic gates provide temporal resolution, periodic gates track
rhythmic structure, and positional gates supply fixed structural
priors. The concurrent LMWT~\cite{kiruluta2025lmwt} applies a similar
multi-resolution principle via learned Haar wavelets; LPA differs in
using explicit rectangular pulses with learned centers and widths.

LPA returns to local processing but makes the receptive field
\emph{data-dependent}. The gate predictor uses a small depthwise
convolution to predict where to look and how wide to integrate.
The depth wall marks where this argument breaks down---wav2vec2's
deep layers perform linguistic computation that requires the
content-dependent global routing attention provides. The SepFormer
cross-validation, where all 16 intra-chunk layers are acoustic and
all are replaceable, is the strongest evidence that the wall reflects
the \emph{nature of computation}, not an architectural ceiling.
Further discussion of theoretical perspectives and open questions
appears in Appendix~\ref{app:theory}.

\section{Limitations}
\label{sec:limitations}

The depth wall at 8--9/12 layers is the primary limitation.
We have shown that it correlates with the acoustic--linguistic
transition, but have not proven that LPA \emph{cannot} perform
linguistic computation. Different gate shapes or training from
scratch might push the wall deeper. The SepFormer cross-validation
supports the hypothesis but retains 16 inter-chunk attention layers
as a confound. All wav2vec2 results are single-run without
confidence intervals, though the consistent depth-wall pattern across
four independent configurations (Table~\ref{tab:progressive})
provides implicit replication. All results use the same
95M-parameter model.
Wav2vec2-large and Whisper are untested. The MLX speedup numbers
(Table~\ref{tab:speed}) combine LPA's algorithmic
$O(n)$ advantage with the framework switch from PyTorch/MPS to MLX;
all speedups use the FP16 attention baseline (see caption). LPA is
complementary to distillation (which reduces model size) and
quantization (which reduces precision). The three axes compose,
and LPA's element-wise operations are particularly amenable to
low-bit quantization.

\section{Conclusion}
\label{sec:conclusion}

We introduced the Learnable Pulse Accumulator, a new $O(n)$
building block for sequence mixing. LPA replaces key-query
matching with learned gating functions (content-dependent
rectangular pulses, periodic Haar-like windows, and
position-dependent basis functions) that define where to
aggregate information, how wide to look, and at what periodicity
to repeat. We developed a practical conversion recipe. An MSE
diagnostic sweep measures per-layer difficulty and determines
replacement order, requiring no architecture search.

At 8/12 replacement, LPA achieves 10.61\% WER on test-clean with
3.27$\times$ measured speedup at 120s audio on M4~Pro vs.\ an FP16
attention baseline, via a dedicated MLX inference path
(1.97$\times$ in the PyTorch/MPS path). The near-binary gate structure at inference enables
dense GPU computation with no CPU--GPU synchronization, and all
operations are compatible with mobile neural accelerators.
Cross-domain validation on SepFormer speech enhancement confirms
the depth wall is domain-specific. All 16 intra-chunk attention
layers are replaced without collapse in a purely acoustic model,
while wav2vec2's linguistic layers resist replacement.

LPA is not a replacement for attention. It is a lower-cost primitive
that works where attention is unnecessary, together with a recipe for
determining where that boundary lies. Combining it with teacher
distillation and hybrid architectures that retain attention only
where needed could close the quality gap and bring efficient
long-form ASR to mobile neural accelerators.

\paragraph{Code availability.}
Code and pretrained checkpoints will be released upon publication.

\clearpage
\appendix

\section{Edge Deployment: Fused Inference Kernels}
\label{sec:edge}

A primary motivation for LPA is on-device speech recognition where
quadratic attention is prohibitive. We describe an optimized inference
path exploiting the structural properties of trained LPA gates.

\subsection{Hard gate inference}

During training, gate temperature anneals from $\tau=3$ (soft) to
$\tau=0.5$ (near-binary). At inference we set $\tau \to 0$, converting
all gate computations to hard thresholds:

\begin{itemize}
\item \textbf{Aperiodic gates.} The softmax over $T$ positions that
  finds each pulse center collapses to argmax.
  The sigmoid product $\sigma\!\bigl(\tfrac{t - s_p}{\tau}\bigr)
  \sigma\!\bigl(\tfrac{e_p - t}{\tau}\bigr)$ (where $s_p = c_p - \delta_p$
  and $e_p = c_p + \delta_p$ are the pulse start and end) becomes an
  indicator $\mathbf{1}[s_p \le t \le e_p]$. Each pulse selects a
  contiguous frame range.

\item \textbf{Periodic gates.} The cosine-threshold gate
  $\sigma\!\bigl(\tfrac{\cos\theta - \cos\pi d}{\tau}\bigr)$
  becomes a step function. On-regions are determined analytically
  from the zero-crossings of $\cos(2\pi t/p - \phi) - \cos(\pi d)$,
  yielding $\lceil T/p \rceil$ contiguous segments per pulse.

\item \textbf{Learned-basis gates.} The sinusoidal basis
  $\sigma(\mathbf{w}^\top \mathbf{f}(t) / \tau)$ thresholds to a
  binary mask, precomputable for any sequence length $T$ since the
  basis weights are content-independent.
\end{itemize}

\noindent This transformation is exact in the limit $\tau \to 0$ and
introduces no additional approximation beyond the training curriculum
that already drives gates toward binary.

\subsection{Prefix-sum accumulation}

With hard gates, the dense accumulation
$\mathbf{a}_p = \sum_t g_{tp} \, \mathbf{x}_t / \sum_t g_{tp}$
reduces to a range-sum. For each contiguous segment $[s_i, e_i]$:

\begin{equation}
\mathbf{a}_p = \frac{1}{|\mathcal{S}_p|} \sum_{i} \bigl(
  \mathbf{C}[e_i] - \mathbf{C}[s_i - 1] \bigr),
  \quad \mathbf{C}[t] = \sum_{t'=0}^{t} \mathbf{x}_{t'}
\label{eq:prefixsum}
\end{equation}

\noindent where $\mathbf{C}$ is the prefix-sum of input features
(computed once in $O(TD)$), and $|\mathcal{S}_p|$ is the total frame
count across all segments. Pulse averages then require only
$O(P)$ indexed reads regardless of segment length.

The broadcast step is similarly sparse. At each position $t$, only
pulses whose segments contain $t$ contribute to the output. In
practice, 3--5 of 12 pulses are active at any given frame.

\subsection{MLX implementation}

We port the full Wav2Vec2 + LPA model to Apple's MLX framework,
computing all gate operations (convolution, MLP, argmax, thresholding)
and accumulation/broadcast as dense GPU operations in fp16 with fp32
accumulation. This eliminates the CPU-GPU synchronization overhead
inherent in PyTorch/MPS's kernel dispatch model, where each LPA layer
requires $\sim$15 separate kernel launches.

The main optimization is computing binary gates as dense tensors
entirely on GPU, then using matrix multiplication for the accumulate
step, $\mathbf{S} = \mathbf{G}^\top \mathbf{X}$, where $\mathbf{G}$
is the $[T \times P]$ binary gate matrix and $\mathbf{X}$ is the
$[T \times D]$ input. Despite $\mathbf{G}$ being binary, the
dense matmul is faster than sparse alternatives at these dimensions
($T \le 6000$, $P = 12$) because it avoids CPU-GPU round-trips for
segment extraction.

For non-LPA encoder layers, we use MLX's fused
\texttt{scaled\_dot\_product\_attention} kernel. The full model runs
in fp16 with fp32 accumulation for numerical stability.

\subsection{Results}

\begin{table}[h]
\centering
\caption{Inference latency on Apple M4 Pro (36\,GB, 273\,GB/s),
  batch=1. 8/12 LPA layers replaced. \emph{Speedup}
  relative to attention baseline. FP16 attention is slower than FP32
  on MPS, so the precision-fair comparison favors LPA even more.}
\label{tab:edge_bench}
\begin{tabular}{lrrrrr}
\toprule
\textbf{Audio} & \textbf{Attn} & \textbf{Attn} & \textbf{LPA (PT)} & \textbf{LPA (MLX)} & \textbf{Speedup} \\
               & \textbf{(fp32)} & \textbf{(fp16)} & \textbf{(fp32)} & \textbf{(fp16)} & \textbf{vs fp16} \\
\midrule
10\,s   &   47.1\,ms &   45.3\,ms &   58.8\,ms &   42.4\,ms & 1.07$\times$ \\
30\,s   &  182.7\,ms &  186.1\,ms &  156.9\,ms &  124.9\,ms & 1.49$\times$ \\
60\,s   &  499.7\,ms &  521.6\,ms &  349.5\,ms &  244.2\,ms & 2.14$\times$ \\
120\,s  & 1668.0\,ms & 1767.1\,ms &  894.8\,ms &  540.1\,ms & 3.27$\times$ \\
\bottomrule
\end{tabular}
\end{table}

\noindent The MLX LPA model achieves 3.27$\times$ speedup over
the FP16 attention baseline at 120\,s and
is 1.66$\times$ faster than the
PyTorch MPS LPA path, confirming that LPA's gate structure is amenable
to framework-level optimization. At 120\,s, the real-time factor is
0.0045, well within real-time constraints for streaming speech
recognition.

The hard-gate path introduces no measurable WER degradation:
evaluation at $\tau=0.01$ vs $\tau \to 0$ produces identical
transcriptions on LibriSpeech dev-clean.

\subsection{Discussion}

The speedup derives primarily from LPA's $O(n)$
gate structure replacing $O(n^2)$ attention, providing increasing
advantage at longer sequences (1.07$\times$ at 10\,s vs 3.27$\times$
at 120\,s). Running attention itself in fp16 on MPS is actually
\emph{slower} than fp32 (Table~\ref{tab:edge_bench}), because the
$n{\times}n$ attention matrix does not benefit from half precision
on Apple Silicon at these sequence lengths. The framework switch from
PyTorch/MPS to MLX contributes fused kernel dispatch, but the
precision change does not inflate the reported speedup.

The structural sparsity of LPA gates (contiguous on-regions separated
by off-regions) is not an incidental property but a direct consequence
of the rectangular pulse parameterization. At inference, this binary
gate structure enables the accumulation to be expressed as a simple
matrix multiplication $\mathbf{G}^\top \mathbf{X}$ with a sparse
binary $\mathbf{G}$, requiring no special sparse kernels at the
dimensions typical of speech ($T \le 7500$, $P = 12$). For longer
sequences where the dense matmul becomes expensive, a prefix-sum
approach (Eq.~\ref{eq:prefixsum}) with a custom Metal kernel can
reduce accumulation to $O(P)$ indexed reads, though at these
dimensions the CPU-GPU synchronization overhead of segment extraction
outweighs the compute savings.

\section{Evaluation Methodology}
\label{app:eval_bug}

Early experiments used batched evaluation with na\"ive
padding. The CTC decoder hallucinated predictions for padded regions,
inflating all WER values by approximately 9\,pp. The corrected
evaluation uses per-sample output length computation to mask padding
before CTC decoding. Training-time evaluation (10s audio filter) yields
3.34\% on dev-clean. Full evaluation (30s filter) yields 3.18\% on
dev-clean and 3.37\% on test-clean, matching the published 3.4\%.

Later experiments use the corrected evaluation natively.
For the Na\"ive column in Table~\ref{tab:progressive} and the top rows
($\dagger$) of Table~\ref{tab:ablation}, we apply a uniform
$-$9.3\,pp offset to the inflated values from the early runs. This
offset is approximate: the baseline correction is 12.67\%$\to$3.34\%
($-$9.33\,pp), and we verified that the corrected 8/12 result
(10.25\%) is within 0.6\,pp of the offset-corrected estimate. These
values should be treated as approximate.

\section{Supplementary Tables}
\label{app:tables}

\begin{table}[ht]
\centering
\caption{Per-layer MSE difficulty from the elastic net diagnostic sweep.
Layers sorted by increasing difficulty. The sweep overprovisions each
layer with 144 pulses (12~$\times$~3~types~$\times$~4~heads) and uses
L1/L2 regularization to prune; the surviving count indicates how much
capacity each layer demands. The MSE ranking determines
replacement order (Sec.~\ref{sec:order_matters}). The per-layer allocation
suggested by the sweep has not been validated end-to-end
(Appendix~\ref{app:theory}).}
\label{tab:mse_difficulty}
\begin{tabular}{@{}rccl@{}}
\toprule
Layer & MSE & Surviving / 144 & Difficulty \\
\midrule
L0  & 0.006 & 48   & easy \\
L2  & 0.007 & 48   & easy \\
L1  & 0.007 & 51   & easy \\
L3  & 0.021 & 57   & moderate \\
L6  & 0.030 & 74   & moderate \\
L5  & 0.034 & 75   & moderate \\
L7  & 0.043 & 87   & hard \\
L8  & 0.046 & 100  & hard \\
L9  & 0.052 & 87   & hard \\
L4  & 0.065 & 89   & hard \\
L10 & 0.096 & 96   & very hard \\
L11 & 0.370 & 107  & very hard \\
\bottomrule
\end{tabular}
\end{table}

\begin{table}[ht]
\centering
\caption{Best WER (\%) during progressive replacement. Columns show
cumulative recipe improvements: training techniques (100h),
architectural additions with 360h data, and deferred replacement
order with MSE auxiliary loss. Na\"ive column estimated via uniform
offset correction (Appendix~\ref{app:eval_bug}).}
\label{tab:progressive}
\begin{tabular}{@{}rcccc@{}}
\toprule
Layers & Na\"ive & +Recipe & +Architecture & +Order \\
\midrule
0  & 3.34  & 3.34  & 3.34  & 3.34 \\
1  & 4.59  & 4.88  & 4.38  & \textbf{4.25} \\
2  & 6.21  & 5.17  & 4.71  & \textbf{4.67} \\
3  & 9.94  & 7.17  & 6.55  & \textbf{4.77} \\
4  & 10.43 & 6.93  & 6.35  & \textbf{5.31} \\
5  & 13.20 & 7.73  & 7.24  & \textbf{5.26} \\
6  & 14.42 & 7.78  & 7.32  & \textbf{5.64} \\
7  & 19.52 & 8.05  & 7.74  & \textbf{7.64} \\
8  & 58.33 & 10.25 & 9.77  & \textbf{9.35} \\
9  & 69.57 & 17.35 & 14.88 & \textbf{14.19} \\
10 & ---   & 37.51 & 28.13 & \textbf{28.02} \\
\bottomrule
\end{tabular}
\end{table}

\begin{table}[ht]
\centering
\caption{WER (\%) across evaluation splits for the +Architecture
configuration. All results use greedy CTC decoding without language
model. Dev-clean baseline differs from Table~\ref{tab:progressive}
(3.18\% vs.\ 3.34\%) because this evaluation includes utterances up to
30s while training-time evaluation filters at 10s.}
\label{tab:test_splits}
\begin{tabular}{@{}rccc@{}}
\toprule
Layers & dev-clean & test-clean & test-other \\
\midrule
0  & 3.18  & 3.37  & 8.67  \\
1  & 4.26  & 4.63  & 11.83 \\
2  & 4.64  & 4.78  & 12.67 \\
4  & 6.38  & 6.62  & 16.52 \\
6  & 7.35  & 7.90  & 18.98 \\
7  & 7.90  & 8.50  & 20.84 \\
8  & 9.92  & 10.61 & 27.10 \\
9  & 15.82 & 16.56 & 37.08 \\
10 & 30.39 & 31.62 & 54.51 \\
\bottomrule
\end{tabular}
\end{table}

\section{Inference Benchmark Details}
\label{app:benchmarks}

Full benchmark results on Apple M4 Pro, including
per-configuration timing and memory estimates.
PyTorch columns use MPS backend in FP32; MLX uses Metal GPU in FP16
with FP32 accumulation.

\begin{table}[ht]
\centering
\caption{Inference time (ms) across configurations and durations on
Apple M4 Pro, batch~1.
PT = PyTorch/MPS. MLX = MLX/Metal (FP16).
Speedup is relative to the attention baseline (FP16).}
\label{tab:speed_full}
\begin{tabular}{@{}rcccccrr@{}}
\toprule
Audio & Base & Base & 8/12 & 12/12 & 8/12 & \multicolumn{2}{c}{Speedup vs Base fp16} \\
      & (fp32) & (fp16) & (fp32) & (fp32) & (MLX) & 12/12 PT & 8/12 MLX \\
\midrule
10s  &   47 &   45 &   59 &   65 &   42 & 0.69$\times$ & 1.07$\times$ \\
30s  &  183 &  186 &  157 &  135 &  125 & 1.38$\times$ & 1.49$\times$ \\
60s  &  500 &  522 &  350 &  246 &  244 & 2.12$\times$ & 2.14$\times$ \\
120s & 1668 & 1767 &  895 &  479 &  540 & 3.69$\times$ & 3.27$\times$ \\
\bottomrule
\end{tabular}
\end{table}

\begin{table}[ht]
\centering
\caption{Per-layer peak memory for gates/attention weights (float32).}
\label{tab:memory}
\begin{tabular}{@{}rcccc@{}}
\toprule
Audio & Frames & Attention & LPA (12p) & Ratio \\
\midrule
10s  &   500 & 0.95\,MB  & 23\,KB  & 42$\times$ \\
30s  & 1,500 & 8.6\,MB   & 70\,KB  & 125$\times$ \\
60s  & 3,000 & 34.3\,MB  & 140\,KB & 250$\times$ \\
120s & 6,000 & 137.3\,MB & 281\,KB & 500$\times$ \\
\bottomrule
\end{tabular}
\end{table}

\section{Roofline Analysis: Pulse Count at Batch Size 1}
\label{app:roofline}

The MSE sweep allocates up to 36 pulses per layer for hard layers
(vs.\ 12 base). A na\"ive FLOP analysis would suggest
3$\times$ slower LPA inference. A roofline analysis on M4 Pro
(273\,GB/s bandwidth, $\sim$16.7 TFLOPS fp16) reveals this
is wrong. At $B{=}1$, the LPA accumulation step is
memory-bandwidth-bound, and the linear projections are near the
compute--bandwidth crossover.

\begin{itemize}[nosep,leftmargin=*]
\item \textbf{Attention} must read/write the $n \times n$ score
matrix per head. At $T{=}6000$ with 12 heads, total score storage is
864\,MB in FP16, the dominant cost.
\item \textbf{LPA accumulation} reads the input tensor
$\mathbf{X} \in \mathbb{R}^{T \times D}$ once (18.4\,MB at $T{=}6000$
in FP32). The gate tensor $\mathbf{G} \in \mathbb{R}^{T \times P}$
adds only $P \times T \times 4$ bytes (864\,KB at $P{=}36$), just 4.7\%
of the input read.
\end{itemize}

\begin{table}[ht]
\centering
\caption{Roofline analysis: per-layer inference time ($\mu$s) on M4 Pro,
$B{=}1$, $T{=}6000$ (120s audio). Four $768{\times}768$ linear
projections dominate LPA time regardless of pulse count.}
\label{tab:roofline}
\begin{tabular}{@{}lccc@{}}
\toprule
Component & Attention & LPA ($P{=}12$) & LPA ($P{=}36$) \\
\midrule
Linear projections & 1,696 & 1,696 & 1,696 \\
$QK^\top$ / gate prediction & 3,311 & 797 & 807 \\
Softmax / element-wise & 2,110 & 201 & 201 \\
$\text{Scores} \times V$ / accumulation & 3,311 & 136 & 142 \\
\midrule
\textbf{Total per layer} & \textbf{10,428} & \textbf{2,835} & \textbf{2,851} \\
\textbf{$\times$ 12 layers} & \textbf{125\,ms} & \textbf{34.0\,ms} & \textbf{34.2\,ms} \\
\bottomrule
\end{tabular}
\end{table}

\noindent Going from 12 to 36 pulses costs $+$16\,$\mu$s per layer
($+$0.2\,ms across 12 layers), a 0.6\% increase.
The four $768{\times}768$ linear projections account for 60\% of LPA
time and are completely pulse-count-independent.

\begin{figure}[ht]
\centering
\includegraphics[width=0.65\textwidth]{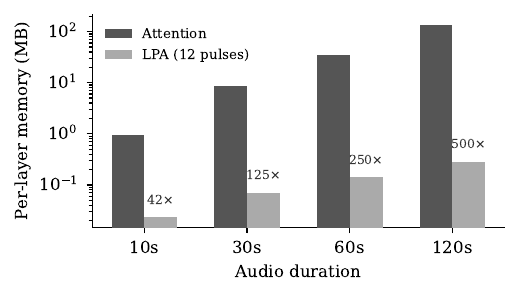}
\caption{Per-layer memory for gates (LPA, $P{=}12$) vs.\ attention
score matrix across audio durations (FP32). LPA memory grows linearly
while attention grows quadratically.}
\label{fig:memory_comparison}
\end{figure}

\section{Theoretical Perspectives}
\label{app:theory}

\subsection{The local--global--learned-local circle}

LPA's intellectual lineage traces a circle in speech recognition
architecture:

\textbf{Local} (1989): Time-Delay Neural Networks used
shared weights across time with fixed local receptive fields.

\textbf{Global} (2017--present): Self-attention provided pairwise
interactions across all positions, enabling arbitrary-range
dependencies, but at $O(n^2)$ cost.

\textbf{Learned-local} (this work): LPA returns to local processing
but makes the receptive field data-dependent. The gate
predictor uses a small depthwise convolution to predict where to look
and how wide to integrate.

\subsection{Open questions}

\textbf{Is the depth wall fundamental?}
We have shown that the wall correlates with the acoustic--linguistic
transition, but have not proven that LPA cannot perform
linguistic computation. Different gate shapes or training from
scratch might push the wall deeper.

\textbf{Fine-tuning confound.}
The SepFormer control experiment (fine-tuning with attention intact:
11.34\,dB vs.\ LPA's 6.82\,dB at 16/16) shows that improvement
over the pretrained baseline is partly driven by continued training.
Disentangling these effects requires matched-budget attention
fine-tuning at each wav2vec2 stage.

\textbf{Training from scratch.}
All experiments convert a pretrained attention model. The pretrained
representations may be biased toward patterns attention can compute
but LPA cannot. Samba-ASR's success training Mamba from scratch
(1.17\% WER, 10K+ hours) suggests this is viable.

\textbf{Per-layer allocation.}
The sweep suggests variable pulse allocation
(Table~\ref{tab:mse_difficulty}), but all progressive results use
fixed allocation. The sweep-derived allocation collapsed at 8/12
due to alignment instability, so its value is unvalidated.

\textbf{Generalization.}
wav2vec2-large (317M, 24 layers) and Whisper are untested. Larger
models may have more redundancy or more distributed representations.

\end{document}